\documentclass[conference]{IEEEtran}
\IEEEoverridecommandlockouts
\usepackage{dblfloatfix}    

\usepackage[belowskip=5pt,aboveskip=0pt]{caption}

\usepackage{hyperref}
\usepackage{cite}
\usepackage{amsmath,amssymb,amsfonts}
\usepackage{algorithmic}
\usepackage{graphicx}
\usepackage{textcomp}
\usepackage{xcolor}
\def\BibTeX{{\rm B\kern-.05em{\sc i\kern-.025em b}\kern-.08em
    T\kern-.1667em\lower.7ex\hbox{E}\kern-.125emX}}
\usepackage{subcaption}

\newcommand{\ep}{{(\varepsilon)}}
\newtheorem{lemma}{Lemma}
\newtheorem{theorem}[lemma]{Theorem}
\def\P{\mathbb{P\,}}
\newcommand{\e}{\text{\textnormal{e}}}
\newcommand{\di}{\text{\textnormal{d}}}
\newcommand{\lo}{\text{\textnormal{o}}}
\newcommand{\I}{\mathbb{I}}
\newcommand{\Obb}{\mathbb{O}}

\newcommand{\halmos}{\vspace{3mm} \hfill \mbox{$\Box$}}

\DeclareMathOperator*{\argmin}{arg\,min}
\newcommand{\THETA}{\theta}
\renewcommand{\tilde}{\widetilde}
\renewcommand{\epsilon}{\varepsilon}
\renewcommand{\varrho}{\rho}

\newcommand{\limitP}{p^\star}
\newcommand{\kopt}{k^\star}
\newcommand{\copt}{c^\star}

\begin{document}

\title{Analyzing large frequency disruptions in power systems using large deviations theory
}

\author{\IEEEauthorblockN{Brendan Patch}
\IEEEauthorblockA{\textit{CWI} \\
Amsterdam, The Netherlands \\
brendan.patch@cwi.nl}
\and
\IEEEauthorblockN{Bert Zwart}
\IEEEauthorblockA{\textit{CWI, IEEE Member} \\
Amsterdam, The Netherlands \\
bert.zwart@cwi.nl}
}

\maketitle

\begin{abstract}
We propose a method for determining the most likely cause, in terms of conventional generator outages and renewable fluctuations, of power system frequency reaching a predetermined level that is deemed unacceptable to the system operator. Our parsimonious model of system frequency incorporates primary and secondary control mechanisms, and supposes that conventional outages occur according to a Poisson process and renewable fluctuations follow a diffusion process. We utilize a large deviations theory based approach that outputs the most likely cause of a large excursion of frequency from its desired level. These results yield the insight that current levels of renewable power generation do not significantly increase system vulnerability in terms of frequency deviations relative to conventional failures. However, for a large range of model parameters it is possible that such vulnerabilities may arise as renewable penetration increases.
\end{abstract}

\begin{IEEEkeywords}
energy systems, power system frequency, renewable energy, stochastic processes, Ornstein--Uhlenbeck process, Poisson process, large deviations theory,  Schilder’s theorem
\end{IEEEkeywords}
\section{Introduction}
It is advocated that a substantial increase in output from wind, water, and solar energy sources is required to combat climate change \cite{gielen2019role}. As a consequence of the highly stochastic nature of the variability in the supply of these renewable energy sources \cite{milan2013turbulent}, their integration into the existing power system is highly challenging.

In particular, the evolution of system frequency may no longer behave as predicted by  traditional deterministic models. Moreover, widespread methods of assessing system stability, where worst case scenarios are considered in terms of combinations of conventional outages, may no longer apply \cite{ahmadyar2017framework}.

Decisions based on these methods often only ensure that systems are able to withstand a single generator failure, but do not ensure robustness to multiple failures occurring within a short amount of time of each other. In the future the combination of a single generator failure and renewable fluctuations may be an issue as well, as power output of a wind farm can drop dramatically within seconds \cite{milan2013turbulent}.

Since system frequency is the primary signal used by control mechanisms in modern power systems, it is crucial to understand how it evolves in a system with a high level of renewable energy penetration. In particular, it is useful to understand where the system's vulnerability is located, which translates into the most likely causes of frequency dips.

This motivates us to consider new methods for determining how significant system frequency dips occur. We consider a parsimonious model of aggregate system frequency that incorporates primary and secondary control mechanisms. We assume that conventional generator outages occur according to a Poisson process (PP) with rate $\lambda$ and renewable generation fluctuates according to an Ornstein--Uhlenbeck (OU) diffusion with diffusion parameter $\sigma$ (our methods generalize to more detailed models). Together, these processes can result in a loss of power generation that causes frequency to trend downwards.

We primarily focus on determining the most likely cause of a significant frequency dip, in terms of conventional outages and renewable fluctuations, as a function of the parameters $\lambda$ and $\sigma$. We consider a setting where such excursions occur infrequently and as such constitute rare events. We propose to use large deviations (LD) theory \cite{dz1998} to formulate and solve an optimization problem. Its solution indicates how many conventional outages and how much renewable fluctuation most likely causes a frequency deviation.

Technically underpinning our work are LD results for diffusions \cite{dz1998}, and a novel scaling of the PP governing conventional outages. The contraction principle is used to map the resulting LD principles into the equation governing power system frequency. The resulting approximation of rare event probabilities are computed using calculus of variations, methods for efficiently evaluating matrix exponentials, and numerical optimization techniques.

Our approach is markedly different from Monte Carlo simulation. In rare event settings, standard simulation methods are unreliable and must be carefully reformulated to be effective (e.g., \cite{moriarty2018frequency}). Such implementations can be difficult to achieve in our dynamic continuous-time setting. Our work complements other studies investigating the most probable failure modes of modern power systems. For example, \cite{chertkov2010predicting} uses instanton theory, and \cite{lee2019robustness} formulates the model as a Lur'e system to simplify analysis. The LD approach that we propose has already been shown to be effective in related studies \cite{nesti2018emergent,nesti2019temperature} and is, to our knowledge, a novel method for studying anomalous fluctuations in power system frequency. Our LD approach gives more explicit answers than the instanton approach, as the associated scaling procedure washes out the details that do not contribute significantly to the rare event of interest.

Our study includes an extensive numerical investigation that demonstrates how our theoretical results can be utilized and provides key qualitative insights. We provide evidence that current levels of renewable power generation do not increase system vulnerability in terms of frequency deviations relative to conventional failures, but that
it is possible that such vulnerabilities may arise as renewable penetration increases.

We also show how our method can be used to examine the influence of model parameters. For example characterizing how perturbing inertia does not cause the probability of system failure to change much for a large range of values of inertia, but that such perturbations can have large effects for some specific values of inertia.

The remainder of this paper is organized as follows. Section~\ref{sec:Model} presents our model of aggregate power system frequency. Section~\ref{sec:Method} provides our key theoretical result, which is used in Section~\ref{sec:Ex} to analyze the model.  Proofs are contained in Section~\ref{sec:Proof}. We conclude in Section~\ref{sec:Out} with an outlook towards future research opportunities.

\section{System Model}\label{sec:Model}
The aggregate power system frequency model we consider incorporates primary and secondary control, and is closely related to the single area models presented in \cite{Vorobev2019}.
The model captures stochastic load dynamics in terms of an OU process $P$, which has mean reversal parameter $\varrho$, diffusion parameter $\sigma$ and stochasticity driven by the zero-mean-unit-variance Brownian motion (BM) $(W(t))_{t\in[0,T]}$, and an independent PP $(N(t))_{t\in[0,T]}$  with rate $\lambda$.
The process $P$ models deviations from net aggregate nominal power generation and consumption in the power system at buses housing stochastic renewable power generators or unpredictable loads.
The process $N$ records the times of conventional power generator failures, allowing the deviation in aggregate power generation from conventional (non-renewable) sources to be quantified.

Throughout this paper, $\dot f$ and $\ddot f$ represent $\di f/\di t$ and $\di^2 f/\di t^2$, and $f^{(i)}$ represents higher order derivatives $\di^i f/\di t^i$.
We assume that a system is operating in equilibrium at time zero and that we are interested in how frequency responds to conventional power disturbances and renewable energy fluctuations.
Therefore, we model deviations from nominal values during an operational interval of $[0,T]$ using the equations
\begin{equation}\label{eq:Model}
\begin{split}
\mu \,\ddot \THETA(t) +\alpha \,\dot \THETA(t)&= P(t)-\delta\, N(t)- \beta\,\THETA(t)\\
\dot P(t) &= -\varrho\,P(t) + \sigma\,\dot W(t)\,,
\end{split}
\end{equation}
where $\dot\THETA$ tracks the per unit frequency deviation of the center of inertia of a power system. We assume $\THETA(0){=}\dot \THETA(0){=}P(0){=}0$. Each conventional failure results in $\delta$ units of power being instantly removed from the system.
The evolution of the system is governed by the inverse aggregate dimensionless droop coefficient $\alpha$, the rate of change of frequency has inertia $\mu$ and is controlled by an integral controller with parameter $\beta$.
These parameters implicitly capture the response of control mechanisms in the system which ramp power up or down depending on frequency and the integral of frequency.

We are interested in the minimum system frequency during the operational interval, known as the {\em nadir}, which is given by the random variable $Z = \min_{t\in[0,T]}\,\dot\THETA(t)$.
Of particular interest is the most likely combination of renewable fluctuations (from their nominal values) and generating unit failures that result in the nadir reaching level $-\gamma$ (where $\gamma$ is a prespecified constant).
The event $\{Z\le -\gamma\}$ depends on the trajectories of the BM $W$ and PP $N$ in a complicated way.
We are interested in the most likely trajectories of renewable fluctuations $P$ and conventional outages $N$ conditional on the event $\{Z\le -\gamma\}$ occurring.

This is related to the instanton approach \cite{chertkov2010predicting}  which effectively maximizes the conditional (on $\{Z\le -\gamma\}$) likelihood over trajectories of $P$ and $N$, after discretizing time. Also of interest is the probability that the nadir reaches a prespecified level $-\gamma$, denoted by $Q(\gamma) = \P\left(Z\le -\gamma\right)$.
In the next section we show how these quantities can be analyzed in a tractable manner using LD theory.
\section{A large-deviations approximation}\label{sec:Method}
This section shows how to obtain an approximation to the most-likely combination of renewable energy fluctuations and conventional generating unit outages associated with large frequency disturbance events $\{Z\le -\gamma\}$.
To that end, consider a scaled version of \eqref{eq:Model}:
\begin{subequations}\label{eq:ApproxModel}
\begin{align}
\mu\,\ddot \THETA^\ep(t) + \alpha\,\dot \THETA^\ep(t) &= P^\ep(t)-\delta\, N^\ep(t)-\beta\,\THETA^\ep(t)\label{eq:ApproxModelA}\\
\dot P^\ep(t) &= -\varrho\,P^\ep(t) + \sqrt{\varepsilon}\tilde \sigma\,\dot W(t)\,,\label{eq:ApproxModelB}
\end{align}
\end{subequations}
where failures occur at the scaled rate $\lambda = \exp(-\tilde\lambda/\varepsilon)$, with $\tilde\sigma$ and $\tilde\lambda$ constants, and $\varepsilon>0$ a scaling parameter,
Let ${\boldsymbol{\theta}}^\ep \equiv (\dot\theta^\ep, \dots, {\theta^{(6)}}^\ep)$ and $Z^\ep = \min_{t\in[0,T]}\,\dot\THETA^\ep(t)$ with corresponding probability of a large frequency disturbance $Q^\ep(\gamma) = \P(Z^\ep \le -\gamma)$.
We will now show how to approximate this probability and identify the most likely combination of renewable energy fluctuations and conventional generating unit outages associated with disturbances.

Given the $7{\times}7$ matrix $\mathcal A$, which we define later, we show that the most relevant limiting trajectories to our study are elements of the family of functions defined by
\begin{equation}\label{eq:SimplematrixExp}
y(t) =  \exp\big(\mathcal A\,t\big) y(0)\,,\quad t\in[0,T]\,,
\end{equation}
with
\[
y(0) = \left[\begin{array}{ccccccc} \Obb_{1,2}& -\frac{k\delta}{\mu} &  c_1 & c_2 & c_3 & -\frac{k\delta a_2}{\mu}+a_3c_2+\frac{\beta\varrho^2\delta k}{\mu^2}\end{array}\right]^{\sf T}.
\]
Here, $k\in\mathbb N$ and $c = (c_1,c_2,c_3) \in\mathbb R^3$ are free variables.  The $i$th coordinate $y_i$ corresponds to ${\theta^{(i-1)}}^\ep$ as $\varepsilon\downarrow 0$.


Given $\mathcal B_1$, $\mathcal B_2, \mathcal B_3$ (defined later in this section), we have the following result, proven in Section V.
\begin{theorem}\label{prop1} As $\varepsilon\downarrow 0$
\[
Q^\ep(\gamma) = \exp\big(- \frac 1\varepsilon J(\copt,\kopt)\big(1+\lo(1)\big)\big)\,,
\]
where $\lo(1) \to 0$ and $J(\copt,\kopt)$ is the optimal value of
\begin{equation}\label{eq:OPTmain}
\begin{array}{cc}
\min\limits_{c\in\mathbb R^3,\,k\in\mathbb N} & J(c,k)\,, \\
\text{s.t.} & \min_{0\le t\le T} y_2(t) \le -\gamma\,,
\end{array}
\end{equation}
where
\begin{equation}\label{eq:RateFunMain}
\begin{split}
J(c,k) &= k\tilde \lambda +\frac{1}{2}\tilde\sigma^{-2} \Big\{y(0)^\top \mathcal B_1 y(0) \\
&\quad\quad\quad\quad\quad+ \mathcal B_2 y(0) + y(0)^\top \mathcal B_3 + (\varrho\delta k)^2T\Big\}\,.
\end{split}
\end{equation}
\end{theorem}
In this result $y_2(t)$ is computed using \eqref{eq:SimplematrixExp}.
The rate of decay of the noise in the scaled model \eqref{eq:ApproxModel}, as given by $J$, provides information on the most likely way a large frequency dip occurs.
Events which are less likely in the original model have a higher rate of decay and consequently become exponentially less likely in the scaled model as $\varepsilon\downarrow0$.
This allows the decay rate to be used to compare the likelihood of different events.
To determine the most likely event, we let $\kopt$ and $\copt$ be the optimizers in \eqref{eq:OPTmain} and the corresponding trajectory of \eqref{eq:SimplematrixExp} be $y^\star$.
The most likely way that a frequency dip occurs is by having $k^*$ generator failures at the beginning of the interval $[0,T]$, and the renewable fluctuations follow the path
\[
\limitP(t) = \mu y^\star_3(t) +\alpha y^\star_2(t) +\beta y^\star_1(t).
\]
We conclude this section by providing expressions for $\mathcal A, \mathcal B_1$, $\mathcal B_2$, and $\mathcal B_3$.
Firstly
\[
\mathcal A = \left[ \begin{array}{cc} \Obb_{7,1} & \begin{array}{c} \I_6 \\ \left[\begin{array}{cccccc} a_1 & 0  & a_2 & 0 & a_3 & 0 \end{array}\right] \end{array} \end{array}\right]\,,
\]
where $\Obb_{i,j}$ is an $i{\times}j$ matrix of zeros, $\I_i$ is an $i{\times}i$ identity matrix, with $a_1 = \mu^{-2}\beta^2\varrho^2$, $a_2  = \mu^{-2}\big(2\beta\varrho^2\mu-\beta^2-\alpha^2\varrho^2\big)$, $a_3 = \mu^{-2}\big(\alpha^2+\varrho^2\mu^2-2\mu\beta\big)$.
Now, define $(\mathcal H)_{ij} = h_ih_j$ with $h_1  = \beta\varrho$, $h_2  = \beta+\varrho\alpha$, $h_3  = \alpha+\varrho\mu$, $h_4  = \mu$, $h_5 = h_6 = h_7=0$, and $h_8=\varrho\delta k$. Let
\[
\mathcal H = \left[\begin{array}{cc} \mathcal H_1 & \mathcal H_2 \\ \mathcal H_2^\top & (\varrho\delta k)^2 \end{array}\right]\,,
\]
where $\mathcal H_1$ and $\mathcal H_2$ are $7{\times}7$ and $7{\times}1$ submatrices.
The matrices $\mathcal B_1$, $\mathcal B_2$, and $\mathcal B_3$ are computed as follows.
Firstly $\mathcal B_1 = \mathcal B_{12}^\top\mathcal B_{11}$, $\mathcal B_3 = \mathcal B_{12}^\top\mathcal B_{31}$ are $7{\times}7$ and $7{\times}1$ matrices with $\mathcal B_2$, $\mathcal B_{11}$, $\mathcal B_{12}$, and $\mathcal B_{31}$ corresponding to $1{\times}7$, $7{\times}7$, $7{\times}7$, and $7{\times}1$ submatrices that follow from evaluating the matrix exponentials
\begin{equation}\notag
\begin{split}
&\e^{\mathcal Q_1 T} = \left[\begin{array}{cc} \mathcal D & \mathcal B_{11} \\ \Obb_{7,7} & \mathcal B_{12} \end{array}\right]\,,\quad \e^{\mathcal Q_2 T} = \left[\begin{array}{cc} \mathcal D & \mathcal B_{2} \\ \Obb_{1,7} & \mathcal D\end{array}\right]\,,\\
&\e^{\mathcal Q_3 T} = \left[\begin{array}{cc} \mathcal D & \mathcal B_{31} \\ \Obb_{1,7}& 0\end{array}\right]\,,
\end{split}
\end{equation}
letting $\mathcal D$ denote entries which are discarded (i.e., are not needed to determine $\mathcal B_1$, $\mathcal B_2$ or $\mathcal B_3$), with
\begin{equation}\notag
\begin{split}
&\mathcal Q_1 = \left[\begin{array}{cc} -\mathcal A^\top & \mathcal H_{1} \\ \Obb_{7,7} & \mathcal A \end{array}\right]\,,\quad
\mathcal Q_2 = \left[\begin{array}{cc} 0 & \mathcal H_{2}^\top \\ \Obb_{7,1} & \mathcal A \end{array}\right]\,, \\
&\mathcal Q_3 = \left[\begin{array}{cc} -\mathcal A^\top & \mathcal H_{2} \\ \Obb_{1,7} &0  \end{array}\right]\,.
\end{split}
\end{equation}

\section{Experimental Results}\label{sec:Ex}
This section contains three illustrative numerical investigations. Unless specified otherwise: $\mu=12$, $\alpha=12.5$, $\beta = 0.05$, and $\varrho =1/30$; these values are in line with the values used in \cite{Vorobev2019}. Additionally, we take $\varepsilon = 0.1$ and $\delta = 1$. This value of $\delta$ corresponds to each conventional failure resulting in $1000$MW being removed from the system, which is the typical loss protected against in Great Britain's power system \cite{techReport2019}. Additionally, $T=2$ minutes (i.e., $120$ seconds).

We utilize the Python 3.7.5 SciPy minimize function using the `SLSQP' routine to find $\copt \in\argmin_c J(c,k)$ for fixed $k$; this procedure is repeated for a range of $k$ values and the best selected.
We report on the resulting $\kopt$ and $\limitP(T)$ values, as described in the previous section.
These are used to quantify the most likely number of generator outages and renewable fluctuation size associated with large frequency disturbances.
The code used to perform these experiments is available at \href{https://github.com/bpatch/power-system-frequency-deviations}{https://github.com/bpatch/power-system-frequency-deviations}.

\subsection{Most likely path to failure in contemporary power systems.}\label{sec:Ex1}
\begin{figure}[!b]
\centering
\hspace{-1mm}
  \begin{subfigure}[b]{.49\linewidth}
    \centering \includegraphics{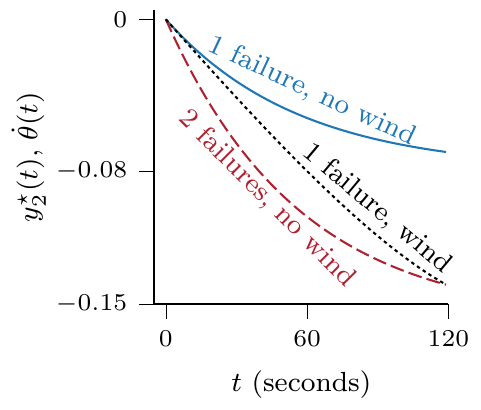}
    \caption{ Frequency deviation.}
    \label{fig:Ex1a}
  \end{subfigure}
   \begin{subfigure}[b]{.49\linewidth}
    \centering \includegraphics{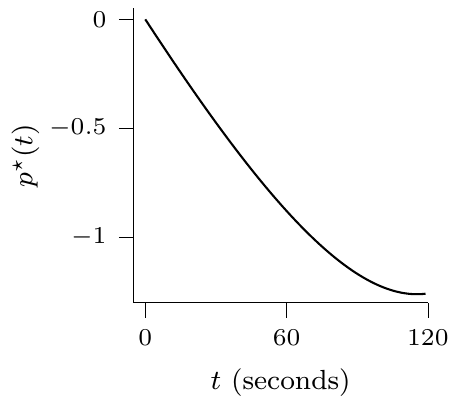}
    \caption{Renewable fluctuation.}
    \label{fig:Ex1b}
   \end{subfigure}
    \caption{(a) Most likely random trajectory to failure (dotted) when $\sigma = 0.2916$ and $\lambda=10^{-3}$ compared with failure scenarios with one (solid) or two (dashed) conventional failures but no renewable fluctuations, and (b) associated renewable fluctuation for random system. }
 \label{fig:Ex1}
\end{figure}

We begin our investigation by considering the effect on the frequency of a system subject to one or two conventional failures which is not subject to renewable fluctuations. In Fig.~\ref{fig:Ex1a} the solid and dashed lines show how the frequency of such a system evolves when one or, respectively, two conventional generating units fail at $t=0$. It can be seen that a single failure results in frequency reaching approximately $-0.0699$Hz within the operational interval and two failures results in the frequency reaching approximately $-0.1397$Hz. The dotted line in this figure is discussed in the next subsection.

In \cite{milan2013turbulent} it is suggested that renewable fluctuations in the range $30{-}50$\% of total capacity over a 2 minute interval at any individual wind park may occur once over several months. For a windfarm such as Hornsea Offshore on the Great Britain power system, which at times produces $737$MW \cite{techReport2019}, this means that a reasonable rare renewable fluctuation is of the order $350$MW and occurs in any two minute operational interval with probability of the order $1/43200\approx0.000023$ (since there are approximately $43200$ two minute operational intervals every two months). Our OU assumption for renewable fluctuations implies $P(2)\sim{\sf Normal}(0,\,\frac{\sigma^2}{2\varrho}(1-\e^{-4\varrho}))$, i.e., approximately $P(2)\sim{\sf Normal}(0,\,\sigma^21.8724)$ given our parameter values. This implies that $\sigma\approx0.0628$ is a reasonable illustrative renewable diffusion coefficient value for current power systems (since this implies $\P(P(2){\le}-0.35){=}1/43200$). After consultation with practitioners, we found it reasonable to illustrate our method under the assumption that approximately two to three conventional generator failures will occur every three days, implying $\lambda = 10^{-3}$ is an appropriate order of magnitude. With these values of $\sigma$ and $\lambda$ our model and analysis suggests that the most likely way for a nadir of $-0.1397$Hz to occur within an arbitrary operational interval is from two conventional generator failures occurring in the operational interval, providing evidence that conventional failures are currently the primary cause of large frequency deviations.

In light of this it is reasonable to ask for what values of $\sigma$ do renewable fluctuations threaten system security? We address this question in the next subsection.

\subsection{Conventional outages vs.\ renewable fluctuations.}\label{sec:Ex2}
\begin{figure}[!b]
  \hspace{-6mm}
  \begin{subfigure}[b]{.48\linewidth}
    \includegraphics{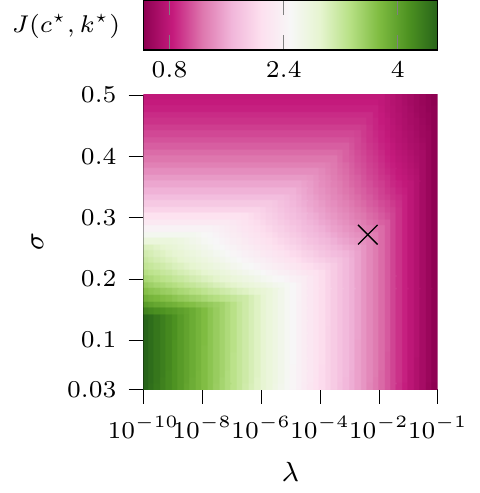}
    \caption{Rate function. }
    \label{fig:Ex2a}
  \end{subfigure}
  \hspace{3mm}
   \begin{subfigure}[b]{.48\linewidth}
    \includegraphics{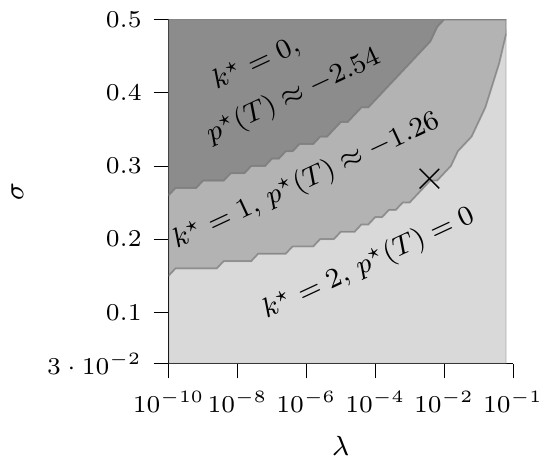}
    \caption{Most likely cause of failure.}
    \label{fig:Ex2b}
   \end{subfigure}
    \caption{(a) Rate function value for a large frequency deviation, and (b) most likely cause of a large frequency deviation. Both as a function of conventional generation failure rate ($\lambda$) and renewable fluctuation intensity ($\sigma$). }
 \label{fig:Ex2}
\end{figure}

We now investigate whether the results from the previous subsection hold for a greater range of renewable fluctuation intensities and conventional failure rates. The plots in Fig.~\ref{fig:Ex2} explore $\sigma\in[0.03,0.5]$ and $\lambda\in[10^{-10}, 10^{-1}]$. To apply Theorem 1, we set $\epsilon = 0.1$. 

Fig.~\ref{fig:Ex2a} provides evidence that the effect of an increase in $\sigma$ or $\lambda$ on the probability of a frequency deviation to $-0.1397$Hz has a strong dependence on the respective value of $\lambda$ or $\sigma$. It can, for example, be seen that that for a high value of either $\sigma$ or $\lambda$ we have $Q^\ep(0.1397)\approx\exp(-0.8/\varepsilon) \approx 0.0003$ regardless of the value of the other parameter. Of note though is that in Fig.~\ref{fig:Ex2b} we see that the cause of the failure probability taking this value is different depending on whether $\sigma$ or $\lambda$ takes on a large value. For high values of $\sigma$ and low values of $\lambda$ the nadir is most likely to be caused by high levels of renewable fluctuations, while for high values of $\lambda$ and low values of $\sigma$ the nadir is most likely to be caused by two conventional failures, but for high values of both parameters the nadir is most likely to be caused by a combination of conventional failures and renewable fluctuations. For any particular smaller value of $\lambda$,  Fig.~\ref{fig:Ex2b} shows that as $\sigma$ increases there are distinct thresholds where renewable fluctuations replace conventional failures as the most likely cause of a large frequency deviation. A key strength of our framework is that it is able to provide system operators with an estimate of where these thresholds are.

Considering $\lambda = 10^{-3}$, as used in the previous subsection, we see that for $\sigma < 0.28$ the most likely cause of a frequency deviation to $-0.1397$ is two conventional generator failures.
At this threshold value of $\sigma$ (marked with a cross) there is a transition in system behavior and the most likely cause of such a frequency deviation becomes a single conventional generator failure accompanied by renewable generation fluctuations. To illustrate this, in Fig.~\ref{fig:Ex1a} the dotted line corresponds to the most likely path to $-0.1397$Hz for a system which is subject to conventional failures at rate $\lambda=10^{-3}$ and renewable fluctuations with variability $\sigma =0.28$.  In Fig.~\ref{fig:Ex1b} the associated most likely renewable fluctuation is $\limitP(T)\approx-1.27$ (i.e, a $1270$MW power loss).

The experiments conducted in this subsection provide further evidence that conventional failures remain to be the primary threat to system security, and that only after substantial increases in renewable penetration would this source of stochasticity present a true threat to system security.

 \subsection{Inertia.}\label{sec:ExParams}
\begin{figure}[!b]
    \centering \includegraphics{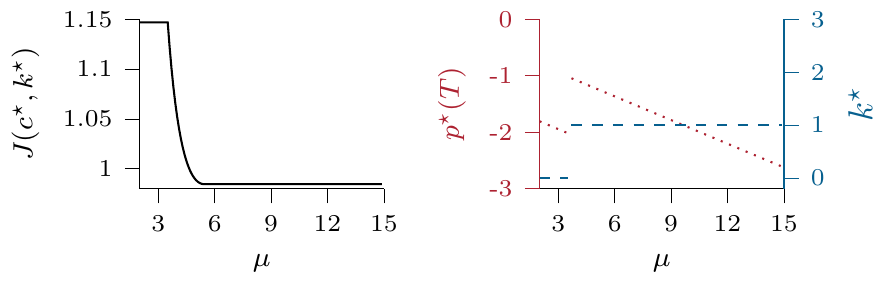}
   \caption{Influence of inertia $\mu$ on LD decay rate $J(\copt,\kopt)$ (left) and the most likely cause ($\kopt$, $\limitP(T)$) (right) with $\gamma=0.1397$, $\lambda=10^{-3}$, $\sigma=0.2916$. }
  \label{fig:Ex3}
 \end{figure}

Inertia influences the likelihood and causes of large deviations.
The curve in the left panel of Fig.~\ref{fig:Ex3} displays the LD decay rate $J(\copt,\kopt)$ as a function of $\mu$.
Recall $Q^\ep(0.1397)\approx \exp(-J(\copt,\kopt)/\varepsilon)$ approximates $Q(0.1397)$, implying that as $J(\copt,\kopt)$ decreases the probability of the frequency deviation increases.
Therefore this curve provides evidence that there are ranged of inertia over which the probability of a large frequency deviation decreases as inertia is increased.
In this example this range is close to a value of inertia where the most likely number of conventional failures increases from zero to one, and the most likely renewable deviation experiences a simultaneous decrease in magnitude from approximately $-2$ to approximately $-1$.
Interestingly there are large ranges of inertia over which changes in inertia do not noticeably affect the probability of a large frequency disruption but where the magnitude of the associated renewable fluctuation increases.

\section{Proof of Main Result}\label{sec:Proof}
To prove Thm.~\ref{prop1} we first provide several lemmas. Our first lemma is a useful integral representation for the frequency in terms of the renewable fluctuations $P$ and outages $N$.

\begin{lemma} Frequency evolves according to
\begin{equation}\label{eq:ExplicitThetaDash}
\dot \theta(t) = \int_0^t a(t-s) [P(s) - \delta N(s)]ds
\end{equation}
with $a(s) = x_3(x_1 e^{x_1 s} - x_2 e^{x_2 s})$, where  $\zeta = \sqrt{\alpha^2-4\beta\mu}$,
\begin{align*}
x_1 = (2\mu)^{-1}(-\alpha+\zeta)\,,\quad x_2 = -(2\mu)^{-1}(\alpha+\zeta)\,,\quad x_3 = \mu/\zeta.
\end{align*}

\end{lemma}
{\bf Proof.} Let $G = [\theta,~\dot\theta]^\top$, then
\[
\di G(t) = \mathcal M G(0)\di t + \left[\begin{array}{c} 0 \\ P(t) - \delta N(t) \end{array}\right]\di t
\]
where
\[
\mathcal M = \left[\begin{array}{cc} 0 & 1 \\ -\beta/\mu & -\alpha/\mu  \end{array}\right]
\]
so
\[
G(t) = \e^{\mathcal M t} G(0) + \int_0^t \e^{ \mathcal M(t-s)} \left[\begin{array}{c} 0 \\ P(s) - \delta N(s) \end{array}\right]\di s\,.
\]
Let $a(s) = (\e^{\mathcal Ms})_{2,2}$. Since $G(0) = [0,~ P(0) - \delta N(0)]^\top$ and $P(0) = 0$, we have that $a(s)$ satisfies \eqref{eq:ExplicitThetaDash} as required.
\begin{raggedright}\halmos\end{raggedright}

We next provide a useful monotonicity result.

\begin{lemma}\label{lem:JumpStart}
If the outage process $N(s)$ is replaced with $\tilde N(s)$ such that $N(s) \leq \tilde N(s)$ for $s\in [0,T]$, then the associated frequency process $\dot{\tilde \theta}(s), s\in [0,T]$ satisfies $\dot\theta(s) \geq \dot{\tilde \theta}(s), s\in [0,T]$. Consequently, given $N(T)=k$, the frequency process $\dot\theta(s)$ is minimized by choosing $N(s) = k$, $s>0$.
\end{lemma}
{\bf Proof.} Note that $x_1>x_2$ in Lem.~\ref{eq:ExplicitThetaDash} implies $a(0) > 0$ and therefore we have $a(s)>0$ for all $s\in[0,T]$. Hence the result follows from \eqref{eq:ExplicitThetaDash}.
\begin{raggedright}\halmos\end{raggedright}

\begin{lemma} Define $\xi_k^\ep(t) = \P(N^\ep(t)= k)$. As $\varepsilon\downarrow 0$,
\[
\varepsilon \log \xi_k^\ep(\nu) \to -\tilde \lambda k
\]
for all $s\in[0,T]$.
\end{lemma}
{\bf Proof.} This follows from
\[
\lim_{\varepsilon\to0} \varepsilon \log\left(\frac{\e^{-k\tilde\lambda  /\varepsilon}T^k}{k!}\e^{-\exp(-\tilde\lambda /\varepsilon)T}\right) = -k\tilde \lambda\,.
\]
\begin{raggedright}\halmos\end{raggedright}

Define the operator $\phi(f) = \inf_{0\le t \le T} \int_0^T a(t-s) f(s) \di s$.
\begin{lemma}
\[
\lim_{\varepsilon\to0}\varepsilon \log \P\left( \inf_{0\le t\le T} \int_0^t a(t-s) P^\ep(s) \di s \le -\eta\right) = -I^\star_g(\eta)\,,
\]
With
$I^\star_g(\eta) = \inf_{\substack{f\in H_0:\\ \phi(f)\le -\eta}} I_d(f)$,
and $I_d(f)$ defined in (7).
\end{lemma}
{\bf Proof.}
Upon removing $N^\ep$ from \eqref{eq:ApproxModel}, $\{\boldsymbol \theta^\ep\}$ satisfies an LDP with good rate function
\begin{equation}\label{eq:NoJumpRateFun}
I_d(f) = \left\{\begin{array}{cc} \frac{1}{2} \int_0^T \mathcal L\big(f, \dot f, \ddot f,   f^{(3)}\big)^2\,\di s\,, & \text{if } f \in H_0\,, \\  \infty\,, & \text{if } f \not\in H_0\,,\end{array}\right.
\end{equation}
where $\mathcal L\big(f, \dot f, \ddot f,   f^{(3)}\big) = \big(\mu f^{(3)} +(\alpha+\varrho\mu)\ddot f+(\beta+\varrho\alpha)\dot f+\beta\varrho f\big)^2$ and
\[
H_x = \{ f \in C_x[0,T] : f(t) = x + \int_0^t g(s) \di s,\, g\in L_2[0,T]\}
\]
is the space of absolutely continuous functions with square integrable derivative.
To see this, from \eqref{eq:ApproxModelB} and \eqref{eq:ApproxModelA},
\begin{align}
\dot W(t) &= \tilde\sigma^{-1}\big( \dot P(t) + \varrho P(t)\big)\,, \label{eq:Wrearrange}\\
P(t) &= \mu \ddot \THETA(t) +\alpha \dot \THETA(t) +\beta \THETA(t)+ \delta k\,,\label{eq:Pr_diff1}\\
\dot P(t) &= \mu  \THETA^{(3)}(t) +\alpha \ddot \THETA(t) +\beta \dot\THETA(t)\,.\label{eq:Pr_diff2}
\end{align}
Upon substituting \eqref{eq:Pr_diff1} and \eqref{eq:Pr_diff2} into \eqref{eq:Wrearrange} and rearranging,
\begin{equation}\notag
\dot W = \tilde\sigma^{-1}\Big(\mu  \THETA^{(3)} +(\alpha+\varrho\mu)\ddot\THETA+(\beta+\varrho\alpha)\dot\THETA+\beta\varrho \THETA+\varrho\delta k\Big)\,.
\end{equation}
Since we removed $N$, we have $k=0$. This characterizes the inverse of a continuous map. Using the contraction principle (see \cite[Thm.\ 4.2.1]{dz1998}) and Schilder's Theorem (see \cite[Thm.\ 5.2.3]{dz1998}), the appropriate rate function for $\{\boldsymbol \theta^\ep\}$ is \eqref{eq:NoJumpRateFun}.
\begin{raggedright}\halmos\end{raggedright}

\begin{lemma}
\[
\lim_{\varepsilon\to0} \varepsilon \log \P\left( N^\ep(T) \ge  k +1 \right) =  (k+1)\tilde \lambda
\]
\end{lemma}
{\bf Proof.}
The lower bound follows from Lemma 4; the upper bound from $\P\left( N^\ep(T) \ge  k +1 \right) \leq e^{-( k +1)\tilde \lambda/\varepsilon}$.
\begin{raggedright}\halmos\end{raggedright}

\begin{lemma}\label{lem:FirstOPT}
As $\varepsilon \downarrow 0$, 
$$
\varepsilon \log Q^\ep(\gamma) \rightarrow - \inf_{k=0,1,\ldots \bar k} [\tilde \lambda k + I^\star_g(\gamma - f(0,k))]
$$
with $f(\nu,k) = \delta k \int_\nu ^T  a(t-s)  \di s$ and $\bar k = \inf \{k: f(0,k) \geq \gamma\}$.

%
%
\end{lemma}
{\bf Proof.}
Note that $f(\nu, k)$ represents the decrease in frequency in $[\nu,T]$ if there have been $k$ generator failures by time $\nu$.
We begin with the asymptotic lower bound. Let
\[
\bar p^\ep(t) = \int_0^t a(t{-}s) P^\ep(s) \di s\,,
\]
then
\begin{align*}
&Q^\ep(\gamma) \ge \P( Z^\ep \le -\gamma\,,\, N^\ep(\nu)\le \bar k)\\
&=\sum_{k=0}^{\bar k} \P( Z^\ep \le -\gamma\,,\, N^\ep(\nu)=k)\\
&\ge \sum_{k=0}^{\bar k} \P\left( \inf_{0\le t \le T} \bar p^\ep(t) {\le} -(\gamma{-}f(\nu,k))\right)\xi_k^\ep(\nu)\,.
\end{align*}
Using Lemma's 4 and 5 and the principle of the largest term, we obtain
\[
\liminf_{\varepsilon \to 0} \varepsilon \log Q^\ep(\gamma)
\geq - \inf_{k\leq \bar k} [\tilde \lambda k + I^\star_g(\gamma - f(\nu,k))]\,.
\]
The lower bound now follows by letting $\nu\downarrow 0$.

For the upper bound, we write
\begin{align*}
Q^\ep(\gamma) \le \P( Z^\ep \le -\gamma\,,\, N^\ep(T)\le \bar k) + \P(N^\ep(T) > \bar k)\,.
\end{align*}
The behavior of the second term is controlled by Lemma 6. Using a similar argument as in the proof of the lower bound,
we can show for the first term
\begin{align*}
& \liminf_{\varepsilon \to 0} \varepsilon \log \P( Z^\ep\le -\gamma\,,\, N^\ep(T)\le \bar k)\\
&\geq - \inf_{k \leq \bar k} [\tilde \lambda k + I^\star_g(\gamma - f(\nu,k))]\,.
\end{align*}
Since $f(\nu, \bar k) \geq \gamma$,
$I^\star_g(\gamma - f(\nu,k))=0$ (no rare-event behavior of $P^{\ep}$ is required) and therefore the first term dominates.
\begin{raggedright}\halmos\end{raggedright}

{\bf Proof of Theorem~\ref{prop1}.}
As the dominant scenario is to have several outages at time $0$,
we can also represent the decay rate of the probability of interest as
\begin{equation}\label{eq:OPT}
\begin{array}{ll}
\min\limits_{f,k} & k\tilde \lambda + I_{d,k}(f)\,, \\
\text{s.t.} & \min_{0\le t\le T} \dot f(t) \le -\gamma\,, \\
& f(0)=\dot f(0)=0\,\ddot f(0) = -k\delta/\mu\,.
\end{array}
\end{equation}
Here, $I_{d,k}(f)$ is the same as $I_d(f)$ defined in (7), except that ${\cal L}$ is replaced by
$\mathcal L\big(f, \dot f, \ddot f,   f^{(3)}\big) = \big(\mu f^{(3)} +(\alpha+\varrho\mu)\ddot f+(\beta+\varrho\alpha)\dot f+\beta\varrho f + \delta k\big)^2$.
Optimality is determined by the Euler-Lagrange equation
\begin{equation}\notag
\nabla_{f} \mathcal L_k -\frac{\di}{\di t}\left(\nabla_{f^{(1)}}\mathcal L_k\right) + \frac{\di^2}{\di t^2}\left(\nabla_{f^{(2)}} \mathcal L_k\right) - \frac{\di^3}{\di t^3}\left(\nabla_{f^{(3)}} \mathcal L_k\right)= 0\,,\
\end{equation}
where $\nabla_f \mathcal L_i$ is the derivative of $\mathcal L_i$ with respect to the function $f$ (see e.g., \cite[p.\ 190]{courant1962methods}). This equation can be simplified to
\begin{equation}\label{eq:OPTcondition}
\begin{split}
&\beta\varrho^2 \delta k-\mu^2 f^{(6)}+\big(\alpha^2+\varrho^2\mu^2-2\mu\beta\big)f^{(4)}\\
&\quad\quad+\big(2\beta\varrho^2\mu-\beta^2-\alpha^2\varrho^2\big)\ddot f + \beta^2\varrho^2 f=0\,.
\end{split}
\end{equation}
Differentiation of \eqref{eq:OPTcondition} leads to
\begin{equation}\label{eq:OPTcondition2}
\begin{split}
&-\mu^2 f^{(7)}+\big(\alpha^2+\varrho^2\mu^2-2\mu\beta\big)f^{(5)}\\
&\quad\quad+\big(2\beta\varrho^2\mu-\beta^2-\alpha^2\varrho^2\big) f^{(3)} + \beta^2\varrho^2 \dot f=0\,,
\end{split}
\end{equation}
and the additional initial condition $f^{(6)}(0) = a_1 f(0) + a_2 \ddot f(0) +a_3 f^{(4)}(0)+\mu^{-2}\beta\varrho^2\delta k$.
Now, for $k=1,\dots,7$ define $y_k = f^{(k-1)}$, then due to \eqref{eq:OPTcondition2}, the solution to \eqref{eq:OPT} is of the form \eqref{eq:SimplematrixExp}, and the function $f$ in (4) is fully specified by $c$ and $k$. In this notation $\mathcal L_k$ becomes $\mathcal L_k\big(y_1, y_2, y_3, y_4\big) =
\big(\mu y_4 +(\alpha+\varrho\mu)y_3+(\beta+\varrho\alpha)y_2+\beta\varrho y_1+\varrho\delta k\big)^2$, so that given our explicit form of $y$. Therefore $J$ in (4) equals
\[
k\tilde\lambda +\frac{1}{2}\tilde\sigma^{-2}\, \left[\begin{array}{cc} y(0)^\top & 1 \end{array}\right]\int_{0}^{T}\e^{\mathcal A^\top t}\,\mathcal H\,\e^{\mathcal At} \di t\, \left[ \begin{array}{c} y(0) \\ 1 \end{array}\right]\,, \\
\]
where $\mathcal H$ is given in Section~\ref{sec:Method}.
The second term in the above expression for $J$ can be rewritten as
\begin{equation}\notag
\begin{split}
\frac{1}{2}\tilde\sigma^{-2} \Big\{ y(0)^\top \int_0^T \e^{\mathcal A t}\mathcal H_1 \e^{\mathcal A t} \di t y(0) +\int_0^T \mathcal H_2^\top \e^{\mathcal A t} \di t y(0) \\
+ y(0)^\top\int_0^T \e^{\mathcal A^\top t}\mathcal H_2 \di t + (\varrho\delta k)^2 T\Big\}\,.
\end{split}
\end{equation}
The three integrals can be evaluated using \cite[Thm.~1]{van1978computing} as
\begin{equation}\notag
\begin{split}
&\int_0^T \e^{\mathcal A t}\mathcal H_1 \e^{\mathcal A t} \di t = \mathcal B_1\,,\quad \int_0^T \mathcal H_2^\top \e^{\mathcal A t} \di t  = \mathcal B_2\,,\\
&\int_0^T \e^{\mathcal A^\top t}\mathcal H_2 \di t  = \mathcal B_3\,,
\end{split}
\end{equation}
where these matrices are defined in Section~\eqref{sec:Method}. Combining these together leads to \eqref{eq:RateFunMain}.
\begin{raggedright}\halmos\end{raggedright}

\section{Outlook}\label{sec:Out}
We have illustrated how LD can be applied to investigate the interplay of conventional outage and renewable fluctuations. We have used a Gaussian process model noise, which might be conservative and therefore the impact of renewable fluctuations compared to conventional generating unit outages may be more significant than our analysis suggests; LD theory can handle more general types of noise. This and several other extensions are called for.
In particular, the influence of the network topology is a key aspect to many frequency deviation questions; early computations of a network version of our model are promising.
We also intend to extend the LD approach to incorporate finite turbine-time constants and deadband controls.
Finally, our analysis can be used to enhance rare event simulation methods, as in \cite{wadman2016large}.

\section*{Acknowledgments}
{\footnotesize This work was supported by NWO grant 639.033.413. We thank Daan Crommelin,  Tommaso Nesti, Andrew Richards, Kostya Turitsyn, Petr Vorobev, and Alessandro Zocca for useful discussions. }

\end{document}